\documentclass[lettersize,journal]{IEEEtran}
\usepackage{amsmath,amsfonts}
\usepackage{algorithmic}
\usepackage{algorithm}
\usepackage{array}
\usepackage[caption=false,font=normalsize,labelfont=sf,textfont=sf]{subfig}
\usepackage{textcomp}
\usepackage{stfloats}
\usepackage{url}
\usepackage{verbatim}
\usepackage{graphicx}
\usepackage{cite}
\usepackage{overpic} 
\usepackage{subcaption} 
\usepackage{booktabs,multirow} 
\usepackage{amssymb} 
\usepackage{bbding} 
\usepackage[numbers]{natbib}
\usepackage[colorlinks,
            linkcolor=red,
            anchorcolor=blue,
            citecolor=green]{hyperref}

\begin{document}

\title{Efficient Star Distillation Attention Network for Lightweight Image Super-Resolution
}


\author{Fangwei Hao, Ji Du, Desheng Kong, Jiesheng Wu, 
Jing Xu, ~\IEEEmembership{Member,~IEEE},  \\ Ping Li, ~\IEEEmembership{Member,~IEEE}



\thanks{Fangwei Hao, Ji Du, Desheng Kong, Jiesheng Wu, and Jing Xu are with the College of Artificial Intelligence, Nankai University, Tianjin, 300350, China.
Ping Li are with the  Department of Computing, Hong Kong Polytechnic University, 999077, Hong Kong, China.
}
\thanks{First Author and Second Author contribute equally to this work.}
}

\markboth{Journal of \LaTeX\ Class Files,~Vol.~14, No.~8, August~2021}%
{Shell \MakeLowercase{\textit{et al.}}: A Sample Article Using IEEEtran.cls for IEEE Journals}


\maketitle

\begin{abstract}
In recent years, the performance of lightweight Single-Image Super-Resolution (SISR) has been improved significantly with the application of Convolutional Neural Networks (CNNs) and Large Kernel Attention (LKA). However, existing information distillation modules for lightweight SISR struggle to map inputs into High-Dimensional Non-Linear (HDNL) feature spaces, limiting their representation learning. And their LKA modules possess restricted ability to capture the multi-shape multi-scale information for long-range dependencies while encountering a quadratic increase in the computational burden with increasing convolutional kernel size of its depth-wise convolutional layer. To address these issues, we firstly propose a Star Distillation Module (SDM) to enhance the discriminative representation learning via information distillation in the HDNL feature spaces. Besides, we present a Multi-shape Multi-scale Large Kernel Attention (MM-LKA) module to learn representative long-range dependencies while incurring low computational and memory footprints, leading to improving the performance of CNN-based self-attention significantly. Integrating SDM and MM-LKA, we develop a Residual Star Distillation Attention Module (RSDAM) and take it as the building block of the proposed efficient Star Distillation Attention Network (SDAN) which possesses high reconstruction efficiency to recover a higher-quality image from the corresponding low-resolution (LR) counterpart. When compared with other lightweight state-of-the-art SISR methods, extensive experiments show that our SDAN with low model complexity yields superior performance quantitatively and visually.
\end{abstract}


\section{Introduction}
Reconstructing a high-resolution (HR) image from corresponding low-resolution (LR) one is the goal of single image super-resolution (SR), a fundamental task in low-level vision. It has drawn the interest of numerous researchers during the last ten years, and significant progress has been made in this field. Many CNN-based methods have been put forth to continuously improve reconstruction performance, since Dong et al. \cite{ref1} first design a three-layer super-resolution convolutional neural network named SRCNN, which achieves surprising performance than the conventional interpolation method \cite{ref2}. A 20-layer network called VDSR is created by Kim et al. \cite{ref3}, and it outperforms SRCNN in terms of effectiveness and demonstrates that one SR model's reconstruction performance can be enhanced by merely deepening the network. Following the residual learning \cite{ref4}, Lim et al. \cite{ref5} develop an improved deep residual network (EDSR) by introducing the residual mechanism into one SR model. Although it results in significant reconstruction performance, the EDSR model is quite heavy. Next, based on residual mechanism and dense connection \cite{ref6},  Zhang et al. develop a residual dense network (RDN) \cite{ref7} and a residual channel attention network (RCAN) \cite{ref8}, both of which are made up of hundreds of convolutional layers. While these are effective SR methods, they are opposed by significant computational loads and massive parameters, hindering their further applications on portable devices with limited computing resource. Therefore, designing lightweight and effective SR networks is essential. 

In terms of lightweight SR, many recent CNN-based methods cope with this task by designing or borrowing advanced architectures or units. Concretely, Tai et al. \cite{ref9} present the deeply-recursive residual network (DRRN), which is derived from the deeply-recursive convolutional network (DRCN) \cite{ref10}. By sharing weights throughout the network, both networks reduce the model parameters. Such a parameter-sharing strategy results in a significant drop in model performance even though it reduces the model complexity. Additionally, other lightweight SR methods \cite{ref11}, \cite{ref12}, \cite{ref13} based on feature distillation strategy are proposed. While the feature distillation strategy is effective for lightweight SR, the reconstruction performance and model efficiency of these distillation SR networks are slightly improved over DRRN, indicating that they are not efficient enough. More recently, PFFN \cite{ref14} adopts pyramid spatial-adaptive feature extraction module (PSAFEM) and the enhanced channel fusion module (ECFM) to balance computational cost and performance, yet it produces restricted performance gain due to the lack of representative long-range dependencies. CFDN \cite{ref15} improves prominent feature distillation network by incorporating convolutional modulation (Conv2Former) with dilated convolution to optimize network performance. However, implementing information distillation in low-dimensional spaces prevents it from achieving discriminative representation learning, resulting in limited reconstruction performance. The similar result is also encountered by FIWHN \cite{ref16} that combines CNN and Transformer to explore a novel architecture which applies wide-residual distillation interaction blocks (WDIBs) with mutual information shuffle and fusion. Besides, HCFormer \cite{ref17} utilizes a hybrid convolution-Transformer architecture which consists of lightweight and efficient convolution blocks (LECB) for potential super-resolution features, and lightweight and efficient Transformer blocks (LETB) for long-term dependency. Nevertheless, they neglect to capture representative contextual information, leading to insufficient reconstruction efficiency. The same limitation goes for the OSFFNet \cite{ref18} which develops an Omni-Stage Feature Fusion (OSFF) architecture to integrate features from different levels and capitalize on their mutual complementarity. 

In addition, attention mechanism \cite{ref19} is a common paradigm used in deep learning, and it is rapidly developed in convolutional neural networks and applied to various computer vision tasks, including image classification \cite{ref20}, \cite{ref21}, \cite{ref22}, \cite{ref23}, \cite{ref24}, \cite{ref25}, object detection \cite{ref26}, \cite{ref27}, \cite{ref28}, \cite{ref29}, semantic segmentation \cite{ref30}, \cite{ref31}, \cite{ref32}, \cite{ref33}, \cite{ref34}, and image restoration \cite{ref35}, \cite{ref36}, \cite{ref37}, \cite{ref38}, \cite{ref39}. Recently, VAN \cite{ref40} develops a large kernel attention (LKA) module by decomposing a large-kernel convolution into a small-kernel depth-wise convolution, a small-kernel depth-wise dilation convolution, and a point-wise convolution. LKA facilitates the learning for long-range dependencies which significantly improve the SR reconstruction performance. However, the computational and memory footprints of depth-wise convolutional layers in LKA rise quadratically as the size of the convolutional kernels increases. The same issue is encountered by LKDN \cite{ref39}, LKASR \cite{ref37} or MSID \cite{ref38} which employ LKA or scalable large kernel attention (SLKA) for lightweight SR task. To alleviate their complexity, Hao et al. \cite{ref41} introduce a large separable kernel attention (LSKA) module \cite{ref25} into lightweight SR task and propose a large coordinate attention network (LCAN). Although their LCAN achieves an effective reconstruction performance with low model complexity, it ignores the learning of multi-shape multi-scale information for long-range dependencies, resulting in limited performance on the public SR datasets.

Overall, existing lightweight image SR methods still show considerable potential for improvement of reconstruction efficiency. In particular, the prominent lightweight information distillation networks are executed in low-dimensional spaces, limiting their SR performance. Besides, existing CNN-based LKA and its improved versions either neglect to learn the multi-shape multi-scale information for long-range dependencies, or incur high computational and memory footprints with increasing convolutional kernel size of its depth-wise convolutional layer.

This paper aims at improving the information distillation ability of CNNs and capturing effective multi-shape multi-scale long-range dependencies, presenting a powerful efficient Star Distillation Attention Network (SDAN) for efficient lightweight SISR. SDAN consists of the proposed Star Distillation Module (SDM) and effective Multi-shape Multi-scale Large Kernel Attention (MM-LKA) module. Specifically, to improve the performance of information distillation modules, SDM is designed to enhance the learning for discriminative representation via information distillation in the high-dimensional non-linear (HDNL) feature spaces. Furthermore, MM-LKA module is developed to learn representative long-range dependencies which further boost the performance of CNN-based self-attention significantly. Additionally, we integrate the proposed SDM, MM-LKA with residual structure \cite{ref4} to generate a residual star distillation attention module (RSDAM) as the building block of our SDAN.

Our contributions are mainly four-fold:
\begin{itemize}
\item{We propose an efficient Star Distillation Attention Network (SDAN) which possesses high reconstruction efficiency to recover a high-resolution (HR) image from the corresponding low-resolution (LR) one for lightweight SR.}

\item{We present a Star Distillation Module (SDM) to enhance the learning for discriminative representation via information distillation in the HDNL feature spaces, and propose a Multi-shape Multi-scale Large Kernel Attention (MM-LKA) module to learn representative long-range dependencies.}

\item{We develop a Residual Star Distillation Attention Module (RSDAM) by integrating the proposed SDM and MM-LKA, and take it as the building block of the proposed SDAN.}

\item{The conducted extensive experiments show that our SDAN with low model complexity achieves superior performance quantitatively and visually, compared with other state-of-the-art lightweight SR methods.}
\end{itemize}

\section{RELATED WORK}
\subsection{Lightweight SR networks}
Deep learning-based image super-resolution has advanced substantially in recent years. The groundbreaking SRCNN, a three-layer convolutional neural network (CNN) which can directly model the mapping function from LR to the corresponding HR, is first proposed by Dong et al. \cite{ref1}. When compared to the earlier interpolation-based method \cite{ref2}, SRCNN demonstrates a significant improvement both quantitatively and visually because of CNN's powerful learning ability. Kim et al. \cite{ref3} develop a very deep super-resolution (VDSR) network with 20 convolutional layers in order to improve performance even more. Next, by adopting the basic recursive structure, Kim et al. \cite{ref10} present a deep recursive convolutional network (DRCN) to ultilize fewer parameters for large receptive field. A deep recursive residual network (DRRN), an enhanced version of DRCN, is later proposed by Tai et al. \cite{ref9}. DRRN performs better than DRCN with the same network depth yet fewer parameters. By integrating the group convolutions and the cascading technique, Ahn et al. \cite{ref45} propose CARN, which makes a trade-off between computation complexity and model performance. Subsequently, Luo et al. \cite{ref39} develop a lightweight SR model called LatticeNet that has comparatively low compute and memory requirements by economically adopting two butterfly structures to combine two residual blocks. Neural architecture search (NAS) technique, which enriches network structures, automatically constructs lightweight SR networks \cite{ref46}, \cite{ref47}. Additionally, CNN-based feature distillation via dimension reduction or channel splitting is another effective strategy. Concretely, Hui et al. \cite{ref11} first apply the feature distillation strategy to the SR task and present a lightweight information distillation network (IDN). IDN possesses quick execution because it uses group convolution and has very few filters per layer. Furthermore, Hui et al. \cite{ref12} improve IDN and propose a lightweight information multi-distillation network (IMDN) by building cascaded information multi-distillation blocks (IMDB), in which the channel splitting strategy is applied multiple times and the channel-wise attention mechanism is introduced. And IMDN takes the first place in the AIM 2019 constrained image SR challenge \cite{ref48} due to its outstanding performance. Then, by combining the proposed feature distillation block with shallow residual connection, Liu et al. \cite{ref13} further develop a residual feature distillation network (RFDN) based on IMDN, and it achieves effective performance while being more lightweight. More recently, in order to balance computational cost and model performance, PFFN \cite{ref14} has adopted the enhanced channel fusion module (ECFM) and pyramid spatial-adaptive feature extraction module (PSAFEM), yet it only yields modest performance gains. To improve SR performance, CFDN \cite{ref15} improves the prominent feature distillation network by incorporating convolutional modulation (Conv2Former) with dilated convolution. Its implementation of information distillation in low-dimensional spaces, however, hinders its efficiency and results in suboptimal reconstruction performance. FIWHN \cite{ref16} employs wide-residual distillation interaction blocks (WDIBs) with mutual information shuffle and fusion for representative learning, and it mixes CNN and Transformer to explore a novel architecture. Besides, a hybrid convolution-Transformer architecture is also used by HCFormer \cite{ref17}, which comprises lightweight and efficient Transformer blocks (LETB) for long-term dependency, and lightweight and efficient convolution blocks (LECB) for potential super-resolution features. In addition, OSFFNet \cite{ref18} develops an Omni-Stage Feature Fusion (OSFF) architecture to integrate features from different levels and capitalize on their mutual complementarity.
\begin{figure*}[!t]
\centering
\includegraphics[width=6.5in]{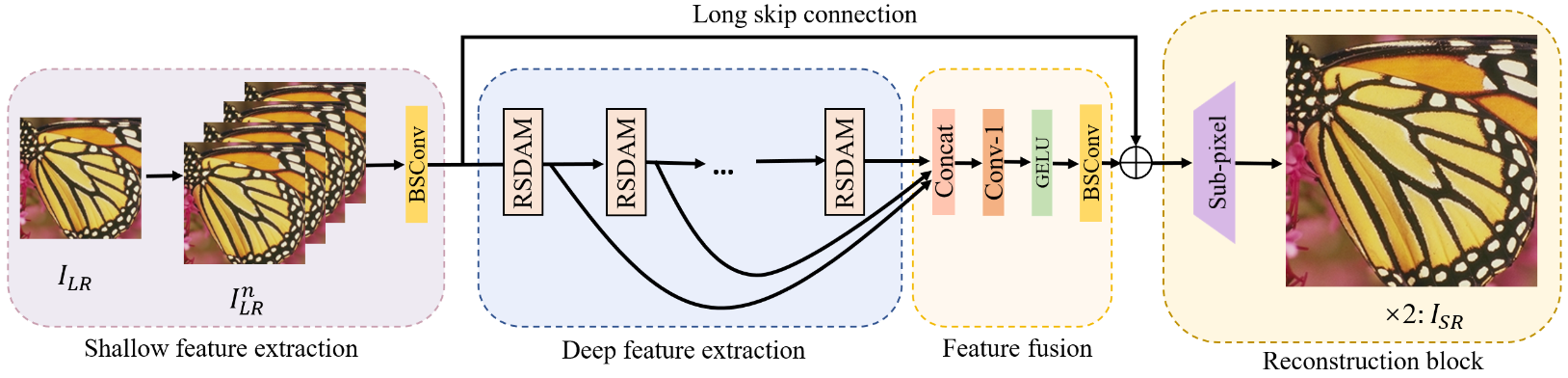}
\caption{Network architecture of our SDAN for ×2 SR.}
\label{fig_1}
\end{figure*}

\subsection{Vision attention}
Viewed as an adaptive reweighting based on its input feature, vision attention has proven to be effective in both low-level tasks (e.g., image SR \cite{ref36}, \cite{ref37}, \cite{ref38}, \cite{ref39}, \cite{ref40}, \cite{ref41}, \cite{ref42}) and high-level tasks (e.g., image classification \cite{ref20}, \cite{ref21}, \cite{ref22}, \cite{ref23}, \cite{ref24}, \cite{ref25}, object detection \cite{ref26}, \cite{ref26}, \cite{ref27}, \cite{ref28}, \cite{ref29}, \cite{ref30}, segmentation \cite{ref31}, \cite{ref32}, \cite{ref33}, \cite{ref34}). The channel attention mechanism is first proposed in \cite{ref20}. After demonstrating its efficacy in image classification, it and its modified versions \cite{ref49}, \cite{ref50} are quickly adopted and utilized in SR networks \cite{ref8}, \cite{ref51}, \cite{ref52}. The channel attention mechanism presents notable gains in reconstruction performance with just refining the feature maps along the channel dimension. Recently, by decomposing a large-kernel convolution into a small-kernel depth-wise convolution, a small-kernel depth-wise dilation convolution, and a point-wise convolution, VAN \cite{ref43} has developed a powerful large-kernel attention (LKA) module. LKA can effectively learn long-range dependencies which greatly enhance the performance of SR reconstruction. Nonetheless, as the size of the convolutional kernel rises, the memory and computing footprints of depth-wise convolutional layers in LKA increase quadratically. LKASR \cite{ref40} develops a scalable large kernel attention (SLKA) which helps to improve the performance of lightweight SR. In order to reduce the complexity of LKA, Lau et al. \cite{ref25} propose a family of large separable kernel attention (LSKA) for high-level tasks (e.g., image classification), by decomposing the two-dimensional (2D) convolutional kernels of the depth-wise convolutional layers in LKA into cascaded one-dimensional (1D) horizontal and vertical kernels, which incurs lower computational complexity. Hao et al. \cite{ref44} further introduce LSKA into lightweight SR task and propose a large coordinate kernel attention network (LCAN) which shows remarkable performance. The success of these LKA-based SR methods indicates the importance of long-range dependencies for reconstruction.

\section{Methodology}
In this section, the overall network architecture of our SDAN is presented at first, followed by the detailed introduction of the proposed star distillation module (SDM). Next, we introduce the multi-shape multi-scale large kernel attention (MM-LKA) module. Subsequently, we provide the details about the residual star distillation attention module (RSDAM). Finally, the loss function is introduced in detail.

\begin{figure*}[!t]
\centering
\includegraphics[width=6in]{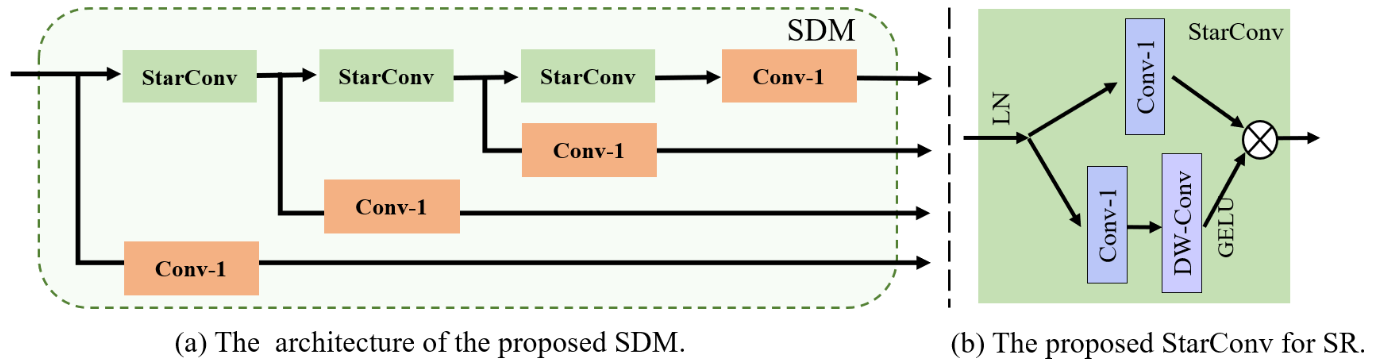}
\caption{The details of the proposed StarConv for high-dimensional mapping and star distillation module (SDM) for lightweight SR.}
\label{fig_2}
\end{figure*}

\subsection{Network architecture}
As Figure \ref{fig_1} shows, similar to previous lightweight SR works \cite{ref13}, our SDAN is primarily composed of four components: the shallow feature extraction part, residual star distillation attention modules (RSDAMs) for deep feature extraction, the feature fusion component, and the reconstruction block. Let $I_{LR}$ represent the input Low-Resolution (LR) image, and $I_{SR}$ denote the output of our SDAN. Initially, we duplicate the input image n times and stack them along the channel dimension. The resulting outcome, denoted as $I_{LR}^n$, serves as the input to the network. Then, we use only one BSConv layer as the shallow feature extraction part to extract the shallow feature $F_{0}$ 
\begin{equation}F_0=H_s(I_{LR}^n),\end{equation}
where $H_{s}(\cdot)$ is the BSConv operation. 
Subsequently, the obtained feature $F_{0}$ undergoes gradual refinement processes through M stacked residual star distillation attention modules (RSDAMs). The refined output feature generated by the last RSDAM represents the deep feature that has been effectively processed during the deep feature extraction part. Consequently, we can express the deep feature extraction process as
\begin{equation}F_{\mathrm{RSDAM}}^k=H_k(F_{\mathrm{RSDAM}}^{k-1}),k=2,\ldots,M,\end{equation}
where $H_k(\cdot)$, $F_{\mathrm{RSDAM}}^{k-1}$, and $F_{\mathrm{RSDAM}}^k$ are the k-th RSDAM operation, its input feature, and its output feature, respectively. After being progressively refined by M RSDAMs, we concatenate the obtained M intermediate features from M RSDAMs along the channel dimension, followed by one 1×1  convolution layer which is utilized to fuse the concatenated features, and a GELU \cite{ref50} function for activation. Then, one 3×3 convolution layer is adopted to smooth the fused features. We can express the process as 
\begin{equation}F_{fused}=H_{fusion}(\mathrm{Concat}(F_{\mathrm{RSDAM}}^1,\ldots,F_{\mathrm{RSDAM}}^M)),\end{equation}
where $\mathsf{Concat}(\cdot)$ represents the concatenation operation, $H_{fusion}(\cdot)$ indicates the function of feature fusion part which consists of a 1×1 convolution layer, a GELU \cite{ref50} activation, and one 3×3 BSConv layer, and $F_{fused}$ denotes the obtained fused feature. Finally, we employ a long skip connection across M RSDAMs to ease the residual learning, and the obtained $F_{fused}+F_0$ is further processed by the reconstruction block which is made up of a 3×3 convolution layer and a sub-pixel convolution layer \cite{ref51} to yield the output SR image $I_{SR}$. The reconstruction block can be formulated as
\begin{equation}I_{SR}=R(F_{fused}+F_0),\end{equation}
where $R\left(\cdot\right)$ is the function of the reconstruction block.
To ensure fair comparison with previous SR methods, such as RFDN \cite{ref13}, FIWHN \cite{ref16}, and OSFFNet \cite{ref18}, we additionally optimize the model using the L1 loss function. Hence, we can formulate the loss function of our SDAN as
\begin{equation}L(\Theta)=\frac{1}{N}\sum_{i=1}^{N}\left\|H_{\mathrm{SDAN}}(I_{LR}^{i})-I_{HR}^{i}\right\|_{1},\end{equation}
Where $H_{\mathrm{SDAN}}\left(\cdot\right)$ and $\mathrm{9}$ denote the function of the proposed SDAN and its learnable parameters, respectively. In addition, Adan \cite{ref53} optimization algorithm, which combines the adaptive optimization, decoupling weight attenuation with modified Nesterov impulse, is adopted to optimize the network.

\begin{figure*}[!t]
\centering
\includegraphics[width=6in]{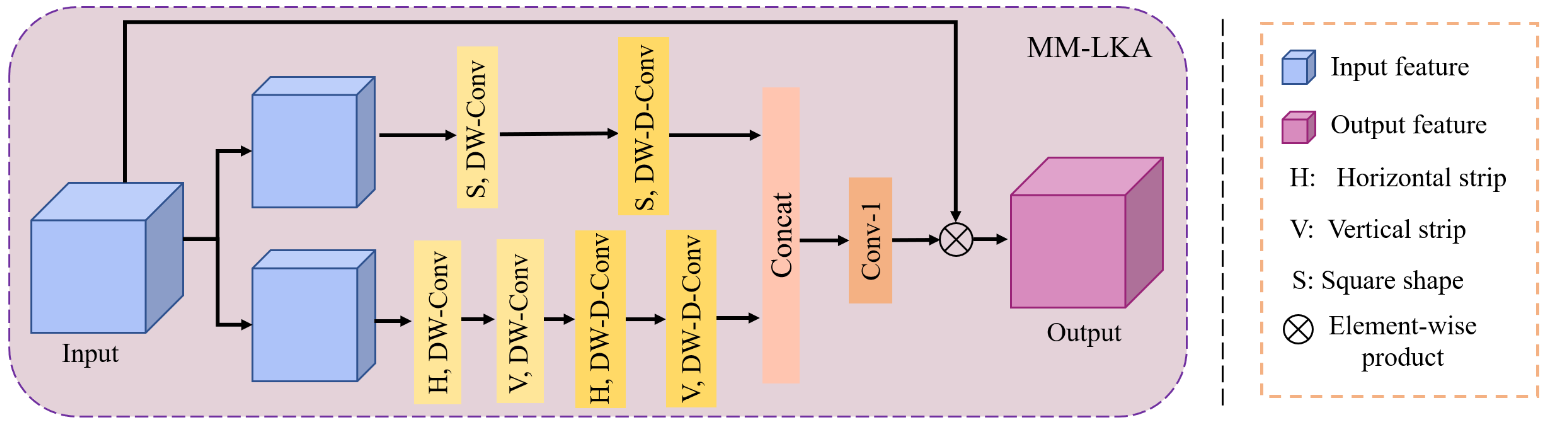}
\caption{The details of the proposed multi-shape multi-scale large kernel attention (MM-LKA).}
\label{fig_3}
\end{figure*}

\subsection{Star Distillation Module (SDM)}

The self-attention mechanism has been widely used in recent years, leading to the element-wise multiplication as a new paradigm for fusing features from different subspaces. We call this paradigm "star operation" for simplicity. In the domains of computer vision (e.g., HorNet \cite{ref54}, VAN \cite{ref40}, FocalNet \cite{ref55}, etc.) and natural language processing (e.g., Monarch Mixer \cite{ref56}, Hyena Hierarchy \cite{ref57}, Mamba \cite{ref58}, etc.), star operation has demonstrated extensive applications and potent performance. StarNet is a straightforward yet effective prototype that Ma et al. \cite{ref59} have proposed more recently. It demonstrates that the star operation can map inputs into HDNL feature spaces. Motivated by this, we propose a star distillation module (SDM) by incorporating the star operation with feature distillation structure based on channel splitting. As Figure \ref{fig_2} shows, SDM allows for information distillation in the HDNL feature spaces, which significantly enhance the learning for discriminative representation for lightweight image super-resolution (SR) task, leading to powerful reconstruction performance.

\subsection{Multi-shape Multi-scale Large Kernel Attention (MM-LKA) module}
We firstly revisit the large kernel attention (LKA) module in \cite{ref40} before introducing the proposed multi-shape multi-scale large kernel attention (MM-LKA) in detail. For LKA, a 13×13 convolution can be decomposed into a 5×5 depth-wise convolution (DW-Conv5), a 5×5 depth-wise dilation convolution with dilation rate 3 (DW-D-Conv5), and a point-wise convolution (Conv1).  Thus, we can formulate the process as
\begin{fontsize}{9pt}{20pt} 
	
\begin{equation}H_{LKA13}(F_{in}^i)=H_{Conv1}\left(H_{DWDConv5}\left(H_{DWConv5}(F_{in}^i)\right)\right)\otimes F_{\mathrm{in}}^i,\end{equation}
\end{fontsize}
where $H_{LKA13}(\cdot)$ denotes the process of LKA, $F_{\mathrm{in}}^i$ is the input feature of i-th LKA module, and $\otimes$ represents element-wise product. Besides, $H_{DWConv5}(\cdot)$, $H_{DWDConv5}(\cdot)$, and $H_{Conv1}(\cdot)$ are the convolutional operations of DW-D-Conv5, DW-Conv5, and Conv1, respectively. We can observe that when the kernel sizes of the depth-wise separable convolutions in LKA increase, its complexity faces a quadratic increase in computational and memory footprints. And it neglects the learning of multi-shape multi-scale information for the long-range dependences.

To deal with the issues, we propose a multi-shape multi-scale large kernel attention (MM-LKA) which consists of one-dimensional (1D) depth-wise convolution, 1D depth-wise dilation convolution, two-dimensional (2D) depth-wise convolution, and 2D depth-wise dilation convolution. The details of the proposed MM-LKA are shown in Figure \ref{fig_3}. Specifically, input features are divided into different groups which are refined respectively by the 1D convolutions and 2D convolutions, followed by a concatenation operation along channels. The concatenated feature maps are then fused by one 1×1 convolution layer to generate self-attention weights to reweight the prior input features $F_{\mathrm{in}}^i$. Thus, MM-LKA possesses the learning ability for multi-shape multi-scale information for the long-range dependences while reducing the computational and memory footprints as the kernel sizes of the depth-wise separable convolutions increase, compared with LKA.

\subsection{Residual star distillation attention module (RSDAM)}
Integrating the proposed SDM and MM-LKA, we propose a residual star distillation attention module (RSDAM) and take it as the building block of the proposed SDAN. RSDAM can not only implement information distillation in HDNL feature spaces but also capture multi-shape multi-scale long-range dependencies, enhancing the discriminative representation learning.

\begin{figure*}[!t]
\centering
\includegraphics[width=6in]{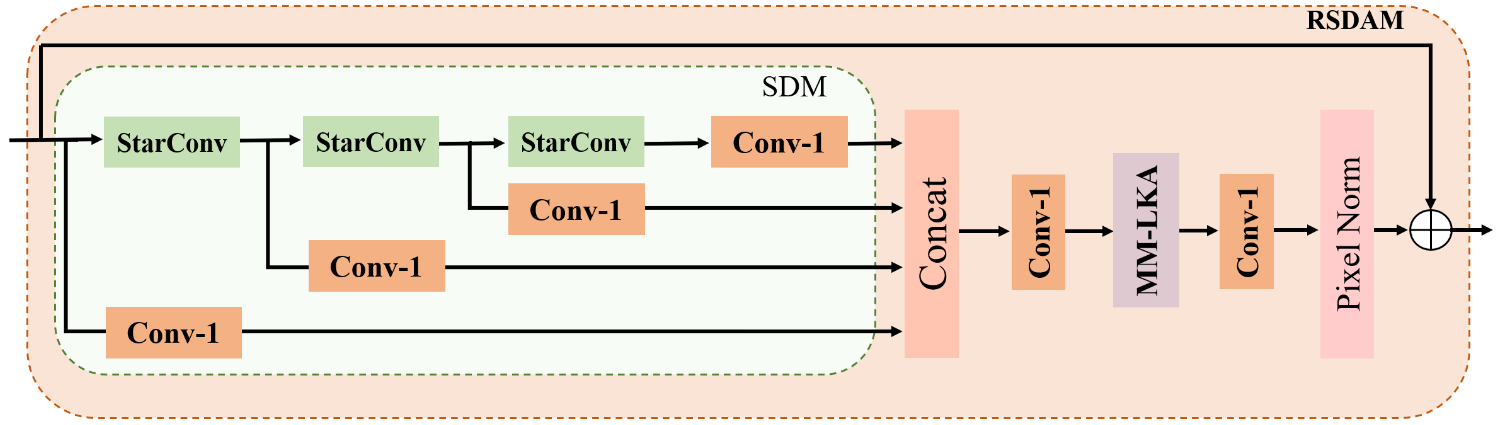}
\caption{The details of the proposed residual star distillation attention module (RSDAM) and its components: the proposed star distillation module (SDM) and the proposed MM-LKA. The Conv-1 and DW-Conv denote 1×1 convolution and depth-wise convolution, respectively.}
\label{fig_4}
\end{figure*}

In the k-th RSDAM, the input feature $F_{\mathrm{RSDAM}}^{k-1}$ is refined gradually. As Figure \ref{fig_4} shows, the process of star distillation module (SDM) can be expressed as 
\begin{equation}\begin{aligned}F_{d_{1}},F_{s_{1}}&=D_1(F_{\mathrm{RSDAM}}^{k-1}),S_1(F_{\mathrm{RSDAM}}^{k-1}),\\F_{d_{2}},F_{s_{2}}&=D_2(F_{d_1}),S_2(F_{s_1}),\\F_{d_{3}},F_{s_{3}}&=D_3(F_{d_2}),S_3(F_{s_2}),\\F_{d_{4}}&=D_4(F_{s_3}),\end{aligned}\end{equation}
where $D_i(\cdot)$ is the i-th 1×1 convolution layer of distillation operation, and $S_{i}(\cdot)$ denotes the corresponding operation of StarConv for feature refinement. $F_{d_i}$ represents the obtained i-th distilled features and $F_{s_i}$ is corresponding refined features.

Subsequently, all of the distilled features generated by 1×1 convolution layers in SDM are concatenated along the channel dimension, and we adopt a 1 × 1 convolution layer to fuse the concatenated features. The process can be expressed as 
\begin{equation}F_f=H_f(\mathrm{Concat}(F_{d_1},F_{d_2},F_{d_3},F_{d_4})),\end{equation}
where $H_{f}(\cdot)$ denotes the function of the 1 × 1 convolution layer, and $F_{f}$ represents the its fused feature. Next, in order to further enhance the discriminative representation learning, we adopt the proposed MM-LKA module to capture multi-shape multi-scale long-range dependencies. We can formulate the process of MM-LKA as
\begin{equation}F_{MM-LKA}=H_{MM-LKA}(F_f),\end{equation}
where $H_{MM-LKA}(\cdot)$, $F_{\mathrm{MM-LKA}}$ represent the function of MM-LKA, its refined features, respectively. Then, we use a 1 × 1 convolution layer to further refine $F_{\mathrm{MM-LKA}}$, with a followed pixel normalization \cite{ref60} module to facilitate stable model training. The process can be formulated as
\begin{equation}F_{\text{normalized}}=\mathrm{Norm}_{\mathrm{pixel}}(Conv_{1\times1}(F_{MM-LKA})),\end{equation}
where $Conv_{1\times1}(\cdot)$, $\mathrm{Norm}_{\mathrm{pixel}}(\cdot)$, and $F_{\text{normalized}}$ represent the process of the 1 × 1 convolution layer, the function of pixel normalization, and the obtained normalized features, respectively. Finally, 
to enable the module to focus on residual learning, we make use of a long skip connection across the module. The process can be formulated as
\begin{equation}F_{\mathrm{RSDAM}}^k=F_{\text{normalized}}+F_{\mathrm{RSDAM}}^{k-1},\end{equation}
where $F_{\mathrm{RSDAM}}^k$ denotes the outcome of the k-th RSDAM.

\section{EXPERIMENTAL RESULTS}
In this section, we firstly the introduce the experimental settings in detail, followed by a series of ablation experiments on SDAN to demonstrate its effectiveness. Next, a qualitative and quantitative comparison of our SDAN with other various lightweight SR methods is presented. Lastly, we conduct the complexity analysis of the proposed SDAN and discuss the limitations.

\subsection{Settings}
To demonstrate the effectiveness and robustness of the proposed SDAN, following \cite{ref18}, we utilize the commonly used DF2K as the training dataset which consists of 800 images from DIV2K \cite{ref61} and 2650 images from Flickr2K \cite{ref5}. For fair comparison with previous SR methods, the corresponding LR images are produced via bicubic downsampling with scaling factors ×2, ×3 and ×4. During training, 64 × 64 patches are generated by random cropping. The test datasets are made up of five common benchmark datasets: SET5 \cite{ref62}, SET14 \cite{ref63}, BSDS100 \cite{ref64}, URBAN100 \cite{ref65}, and MANGA109 \cite{ref66}. The output results of the SR model are transformed to YCbCr space to calculate the peak signal-to-noise ratio (PSNR) and structural similarity index Measure (SSIM) \cite{ref67} to quantitatively evaluate the model performance.
Following \cite{ref39}, we optimize our SR model from scratch using the ADAN \cite{ref53} optimizer with $\beta_1=0.98$, $\beta_2=0.92$ and $\beta_{3}=0.99$. The exponential moving average (EMA) is set to 0.999 to facilitate training. All experiments are conducted using the Pytorch \cite{ref68} framework with one NVIDIA 3090 GPU. The learning rate is set to a constant $5\times10^{-3}$ to train the network for $1\times\dot{10^6}$ iterations.

\subsection{Ablation studies}
\textbf{Break-down ablation.} To evaluate the effectiveness of the proposed SDAN and the efficiency of its components, i.e., SDM and MM-LKA, we carry out a series of ablation experiments.

Initially, we use the DF2K dataset for ×4 SR to train the model without the SDM and MM-LKA, using the tested performance of 32.21 dB PSNR on the Set5 (×4) dataset as the baseline. 

The model with the SDM only is then trained and tested using the same datasets. The SDM has a strong feature extraction capability and can enhance reconstruction performance by capturing discriminative features via information distillation in the HDNL feature spaces, as verified by the test result which increases by 0.12 dB to 32.33 dB PSNR. 

In order to validate the effectiveness of the MM-LKA, we repeat the same experiment on the model with MM-LKA only. The test result rises by 0.22 dB to 32.43 dB in comparison to the baseline. Notably, MM-LKA can substantially improve the performance of the benchmark model in comparison to SDM.

Finally, the model with both SDM and MM-LKA is trained and tested under the same experimental settings to confirm the overall performance of SDAN. The obtained result reaches 32.54 dB, demonstrating the impressive performance of the suggested SDAN for lightweight SR. Table \ref{tab_1} shows all of the experimental results.

\begin{table}[htbp]
  \centering
  \caption{Ablation results from scratch on the Set5 dataset for ×4 SR. The best result is \textbf{highlighted}.}
    \begin{tabular}{p{5.045em}p{3.045em}p{3.045em}p{3.045em}p{3.045em}}
    \toprule
    Baseline & \multicolumn{1}{p{4.045em}}{\Checkmark} & \multicolumn{1}{p{3.045em}}{\Checkmark} & \multicolumn{1}{p{3.045em}}{\Checkmark} & \multicolumn{1}{p{3.045em}}{\Checkmark} \\
    SDM   & \multicolumn{1}{p{4.045em}}{\XSolidBrush} & \multicolumn{1}{p{3.045em}}{\Checkmark} & \multicolumn{1}{p{3.045em}}{\XSolidBrush} & \multicolumn{1}{p{3.045em}}{\Checkmark} \\
    MM-LKA & \multicolumn{1}{p{4.045em}}{\XSolidBrush} & \multicolumn{1}{p{3.045em}}{\XSolidBrush} & \multicolumn{1}{p{3.045em}}{\Checkmark} & \multicolumn{1}{p{3.045em}}{\Checkmark} \\
    \midrule
    PSNR/dB & 32.21 & 32.33 & 32.43 & \textbf{32.54} \\
    \bottomrule
    \end{tabular}%
  \label{tab_1}%
\end{table}%

\textbf{Ablation of SDM.} {In order to validate the efficiency of the proposed star distillation module (SDM), we conduct corresponding experiments for ×3 SR to show the selection of kernel size of depth-wise convolution in StarConv in SDM. In particular, the depth-wise convolutions in StarConv are separately implemented with different kernel sizes, ranging from 3 to 7. Considering the module complexity, we do not choose larger kernel sizes. Subsequently, under the same dataset and experimental setting, we carry out corresponding experiments to train the models, and test the trained models on Manga109 \cite{ref66} dataset for ×3 SR. The obtained experimental results are presented in Table \ref{tab_2}. It can be seen that the performance of the model improves as the kernel size increases, demonstrating the effectiveness of the proposed SDM.}

\begin{table}[htbp]
  \centering
  \caption{Ablation results of SDM on the Manga109 dataset for ×3 SR. The best result is \textbf{highlighted}.}
    \begin{tabular}{p{9.045em}ccc}
    \toprule
    kernel size (kz) & 3     & 5     & 7 \\
    \midrule
    \textcolor[rgb]{ .137,  .122,  .125}{Network Params (K)} & 405   & {419} & {451} \\
    PSNR (dB) & {34.16} & {34.20} & \textbf{34.25} \\
    \bottomrule
    \end{tabular}%
  \label{tab_2}%
\end{table}%

\textbf{Ablation of MM-LKA.} To exploit the effect of different kernel sizes of square and strip depth-wise convolutions in the proposed multi-shape multi-scale large kernel attention (MM-LKA), convolutional kernels with different sizes are adopted. Concretely, the kernel sizes of square depth-wise convolution and square dilated depth-wise convolution in MM-LKA vary from 5×5 to 7×7, and those of strip depth-wise convolution and strip dilated depth-wise convolution are set to 7 or 11 to capture more contextual information. Then, we separately conduct corresponding experiments under the same training dataset and experimental setting for ×4 SR, and we test the trained models on the Set5 (×4) dataset. Table \ref{tab_3} shows the detailed experimental results. We can see that as the square convolution kernel increases, the model achieves certain performance improvement, and the same trend goes for the strip depth-wise convolution. Moreover, models with both square depth-wise convolution and strip depth-wise convolution produce higher performance than single-shape convolution. Among them, the model which possesses 7×7 square convolution and strip depth-wise convolution with size of 11, yields the highest 32.54 dB PSNR. These experimental results indicate that the multi-shape multi-scale contextual information is crucial for efficient lightweight SR task. 
\begin{table}[htbp]
  \centering
  \caption{Ablation results of the proposed MM-LKA on the Set5 dataset for ×4 SR. The best result is \textbf{highlighted}.}
    \begin{tabular}{p{4.045em}p{2.045em}p{2.045em}p{2.045em}p{2.045em}p{2.045em}p{2.045em}}
    \toprule
    \multirow{2}[0]{*}{square size} & 5×5   & \multicolumn{1}{p{2.045em}}{\Checkmark} & \multicolumn{1}{p{2.045em}}{\XSolidBrush} & \multicolumn{1}{p{2.045em}}{\XSolidBrush} & \multicolumn{1}{p{2.045em}}{\XSolidBrush} & \multicolumn{1}{p{2.045em}}{\XSolidBrush} \\
    \multicolumn{1}{r}{} &  7×7  & \multicolumn{1}{p{2.045em}}{\XSolidBrush} & \multicolumn{1}{p{2.045em}}{\Checkmark} & \multicolumn{1}{p{2.045em}}{\Checkmark} & \multicolumn{1}{p{2.045em}}{\Checkmark} & \multicolumn{1}{p{2.045em}}{\XSolidBrush} \\
    \midrule
    \multirow{2}[0]{*}{strip size k} & \multicolumn{1}{l}{7} & \multicolumn{1}{p{2.045em}}{\Checkmark} & \multicolumn{1}{p{2.045em}}{\Checkmark} & \multicolumn{1}{p{2.045em}}{\XSolidBrush} & \multicolumn{1}{p{2.045em}}{\XSolidBrush} & \multicolumn{1}{p{2.045em}}{\XSolidBrush} \\
    \multicolumn{1}{c}{} & \multicolumn{1}{l}{11} & \multicolumn{1}{p{2.045em}}{\XSolidBrush} & \multicolumn{1}{p{2.045em}}{\XSolidBrush} & \multicolumn{1}{p{2.045em}}{\Checkmark} & \multicolumn{1}{p{2.045em}}{\XSolidBrush} & \multicolumn{1}{p{2.045em}}{\Checkmark} \\
    \midrule
    Params/K & - & 393   & 404   & 410   & 419   & 396 \\
    PSNR/dB & - & 32.45 & 32.49 & \textbf{32.54} & 32.47 & 32.44 \\
    \bottomrule
    \end{tabular}%
  \label{tab_3}%
\end{table}%

\begin{table}[htbp]
  \footnotesize 
  \centering
  \caption{Quantitative comparison of our SDAN with the heavy SR methods on benchmark datasets for ×4 SR.}
    \begin{tabular}
    {p{5.345em}p{2.045em}p{2.745em}p{2.745em}p{2.745em}p{3.545em}}
    \toprule
    Method & Param & Set5  & Set14 & BSD100 & Urban100 \\
    \midrule
    EDSR  & 43.1M & 32.46dB & 28.80dB & 27.70dB & 26.64dB \\
    RDN   & 22.3M & 32.47dB & 28.80dB & 27.72dB & 26.60dB \\
    RCAN  & 15.6M & 32.63dB & 28.87dB & 27.77dB & 26.82dB \\
    DRN   & 9.8M  & 32.74dB & 28.98dB & 27.83dB & 27.03dB \\
    ERAN  & 8.02M & 32.66dB & 28.92dB & 27.79dB & 26.86dB \\
    SDAN/ours & 0.41M & 32.54dB & 28.81dB & 27.72dB & 26.65dB \\
    \bottomrule
    \end{tabular}%
  \label{tab_5}%
\end{table}%

\subsection{Comparisons with other SR Methods}
The effectiveness of our SDAN is further verified by quantitatively and visually comparing our experimental results for upscaling factors of ×2, ×3, and ×4 with the results of other previous state-of-the-art lightweight SR networks, including SRCNN \cite{ref1}, VDSR \cite{ref3}, DRCN \cite{ref10}, DRRN \cite{ref9}, IDN \cite{ref11}, IMDN \cite{ref12}, RFDN \cite{ref13}, PFFN \cite{ref14}, CFDN \cite{ref15}, DIIN \cite{ref15_1}, FIWHN \cite{ref16}, EConvMixN \cite{ref69}, and OSFFNet \cite{ref18}.

\begin{table*}[htbp]
  \centering
  \caption{Quantitative results of our SDAN and previous state-of-the-art lightweight SR methods on benchmark datasets. The best and second-best results are \textbf{highlighted} and \underline{underlined}, respectively.}
    \begin{tabular}{p{4.745em}p{3.545em}cp{4.5em}p{5.68em}p{5.09em}p{5.365em}p{5.145em}p{5.455em}}
    \toprule
    \multirow{2}[2]{*}{Method} & \multirow{2}[2]{*}{Years} & \multicolumn{1}{r}{\multirow{2}[2]{*}{Scale}} & \multirow{2}[2]{*}{Params} & Set5  & Set14 & BSD100 & Urban100 & Manga109 \\
    \multicolumn{1}{r}{} & \multicolumn{1}{c}{} &       & \multicolumn{1}{r}{} & PSNR/SSIM & PSNR/SSIM & PSNR/SSIM & PSNR/SSIM & PSNR/SSIM \\
    \midrule
    Bicubic  & -     & \multicolumn{1}{c}{\multirow{14}[2]{*}{×2}} & -     & 33.66/0.9299 & 30.24/0.8688 & 29.56/0.8431 & 26.88/0.8403 & 30.80/0.9339 \\
    SRCNN  & TPAMI15 &       & 8K    & 36.66/0.9542 & 32.45/0.9067 & 31.36/0.8879 & 29.50/0.8946 & 35.60/0.9663 \\
    VDSR  & CVPR16 &       & 666K  & 37.53/0.9587 & 33.03/0.9124 & 31.90/0.8960 & 30.76/0.9140 & 37.22/0.9750 \\
    DRCN  & CVPR16 &       & 1774K & 37.63/0.9588 & 33.04/0.9118 & 31.85/0.8942 & 30.75/0.9133 & 37.55/0.9732 \\
    DRRN  & CVPR17 &       & 298 K & 37.74/0.9591 & 33.23/0.9136 & 32.05/0.8973 & 31.23/0.9188 & 37.88/0.9749 \\
    IDN   & CVPR18 &       & 553K  & 37.83/0.9600 & 33.30/0.9148 & 32.08/0.8985 & 31.27/0.9196 & 38.01/0.9749 \\
    IMDN  & MM19  &       & 694K  & 38.00/0.9605 & 33.63/0.9177 & 32.19/0.8996 & 32.17/0.9283 & 38.88/0.9774 \\
    RFDN  & ECCV20 &       & 534K  & 38.05/0.9606 & 33.68/0.9184 & 32.16/0.8994 & 32.12/0.9278 & 38.88/0.9773 \\
    PFFN  & SPL24 &       & 572 K & \underline{38.14}/0.9608 & 33.76/0.9190 & 32.29/\underline{0.9009} & 32.48/0.9317 & -/- \\
    CFDN  & ICME24 &       & 820 K & 38.11/\textbf{0.9613} & \underline{33.82}/\underline{0.9199} & 32.26/0.9008 & 32.62/0.9328 & -/- \\

    EConvMixN & EAAI24 &       & 953 K & 38.12/0.9609 & 33.70/0.9189 & 32.25/0.9007 & 32.44/0.9317 & 39.06/0.9778 \\
    OSFFNet & AAAI24 &       & 516K  & 38.11/0.9610 & 33.72/0.9190 & \textbf{32.29}/\textbf{0.9012} & \underline{32.67}/\underline{0.9331} & \underline{39.09}/0.9780 \\
    DIIN & TIM24 &       & 726K  & 38.06/0.9610 & 33.73/0.9189 & 32.20/{0.8998} & {32.37}/{0.9301} & {38.95}/0.9778 \\
    FIWHN  & TMM24 &       & 705K  & \textbf{38.16}/\textbf{0.9613} & 33.73/0.9194 & \underline{32.27}/0.9007 & 32.75/0.9337 & 39.07/\underline{0.9782} \\
    SDAN  & ours  &       & 405 K &  \underline{38.14}/\underline{0.9611} & \textbf{33.95}/\textbf{0.9203} & \textbf{32.29}/\textbf{0.9012} & \textbf{32.82}/\textbf{0.9344} & \textbf{39.21}/\textbf{0.9783} \\
    \midrule
    Bicubic & -     & \multicolumn{1}{c}{\multirow{13}[2]{*}{×3}} & -     & 30.39/0.8682 & 27.55/0.7742 & 27.21/0.7385 & 24.46/0.7349 & 26.95/0.8556 \\
    SRCNN  & TPAMI15 &       & 8K    & 32.75/0.9090 & 29.30/0.8215 & 28.41/0.7863 & 26.24/0.7989 & 30.48/0.9117 \\
    VDSR  & CVPR16 &       & 666K  & 33.66/0.9213 & 29.77/0.8314 & 28.82/0.7976 & 27.14/0.8279 & 32.01/0.9340 \\
    DRCN  & CVPR16 &       & 1774K & 33.82/0.9226 & 29.76/0.8311 & 28.80/0.7963 & 27.15/0.8276 & 32.24/0.9343 \\
    DRRN  & CVPR17 &       & 298 K & 34.03/0.9244 & 29.96/0.8349 & 28.95/0.8004 & 27.53/0.8378 & 32.71/0.9379 \\
    IDN   & CVPR18 &       & 553K  & 34.11/0.9253 & 29.99/0.8354 & 28.95/0.8013 & 27.42/0.8359 & 32.71/0.9381 \\
    IMDN  & MM19  &       & 703K  & 34.36/0.9270 & 30.32/0.8417 & 29.09/0.8046 & 28.17/0.8519 & 33.61/0.9445 \\
    RFDN  & ECCV20 &       & 541K  & 34.41/0.9273 & 30.34/0.8420 & 29.09/0.8042 & 28.21/0.8525 & 33.67/0.9449 \\
    CFDN  & ICME24 &       & 828 K & 34.50/0.9286 & 30.41/\textbf{0.8459} & 29.20/\underline{0.8081} & \underline{28.50}/0.8587 & -/- \\

    EConvMixN & EAAI24 &       & 965 K & 34.56/0.9286 & 30.48/0.8452 & 29.19/0.8077 & 28.43/0.8586 & 33.92/0.9467 \\
    OSFFNet & AAAI24 &       & 524 K & \underline{34.58}/\underline{0.9287} & 30.48/0.8450 & \underline{29.21}/0.8080 & 28.49/0.8595 & \underline{34.00}/\underline{0.9472} \\
    DIIN & TIM24 &       & 735K  & 34.48/0.9280 & 30.44/0.8436 & 29.13/{0.8062} & {28.35}/{0.8551} & {33.87}/0.9461 \\
    FIWHN  & TMM24 &       & 713 K & 34.50/0.9283 & \underline{30.50}/0.8451 & 29.19/0.8077 & \textbf{28.62}/\underline{0.8607} & 33.97/\underline{0.9472} \\
    SDAN  & ours  &       & 408 K & \textbf{34.60}/\textbf{0.9289} & \textbf{30.52}/\underline{0.8456} & \textbf{29.22}/\textbf{0.8085} & \textbf{28.62}/\textbf{0.8624} & \textbf{34.16}/\textbf{0.9479} \\
    \midrule
    Bicubic  & -     & \multicolumn{1}{c}{\multirow{14}[2]{*}{×4}} & -     & 28.42/0.8104 & 26.00/0.7027 & 25.96/0.6675 & 23.14/0.6577 & 24.89/0.7866 \\
    SRCNN  & TPAMI15 &       & 8K    & 30.48/0.8626 & 27.50/0.7513 & 26.90/0.7101 & 24.52/0.7221 & 27.58/0.8555 \\
    VDSR  & CVPR16 &       & 666K  & 31.35/0.8838 & 28.01/0.7674 & 27.29/0.7251 & 25.18/0.7524 & 28.83/0.8870 \\
    DRCN  & CVPR16 &       & 1774K & 31.53/0.8854 & 28.02/0.7670 & 27.23/0.7233 & 25.14/0.7510 & 28.93/0.8854 \\
    DRRN  & CVPR17 &       & 298 K & 31.68/0.8888 & 28.21/0.7720 & 27.38/0.7284 & 25.44/0.7638 & 29.45/0.8946 \\
    IDN   & CVPR18 &       & 553K  & 31.82/0.8903 & 28.25/0.7730 & 27.41/0.7297 & 25.41/0.7632 & 29.41/0.8942 \\
    IMDN  & MM19  &       & 715K  & 32.21/0.8948 & 28.58/0.7811 & 27.56/0.7353 & 26.04/0.7838 & 30.45/0.9075 \\
    RFDN  & ECCV20 &       & 550K  & 32.24/0.8952 & 28.61/0.7819 & 27.57/0.7360 & 26.11/0.7858 & 30.58/0.9089 \\
    PFFN  & SPL24 &       & 572 K & 32.35/0.8963 & \underline{28.77}/0.7837 & \underline{27.72}/0.7390 & 26.29/0.7903 & -/- \\
    CFDN  & ICME24 &       & 838 K & 32.33/\underline{0.8981} & 28.65/\underline{0.7855} & {27.70}/\underline{0.7407} & 26.39/0.7957 & -/- \\
    EConvMixN & EAAI24 &       & 983 K & 32.35/0.8972 & 28.69/0.7844 & 27.66/0.7398 & 26.30/0.7943 & 30.76/0.9125 \\
    OSFFNet & AAAI24 &       & 537K  & \underline{32.39}/0.8976 & 28.75/0.7852 & 27.66/0.7393 & 26.36/0.7950 & 30.84/0.9125 \\
    DIIN & TIM24 &       & 747K  & 32.35/0.8963 & 28.73/0.7842 & 27.63/{0.7378} & {26.35}/{0.7920} & {30.81}/0.9119 \\
    FIWHN  & TMM24 &       & 725K  & 32.30/0.8967 & 28.76/0.7849 & 27.68/0.7400 & \underline{26.57}/\underline{0.7989} & \underline{30.93}/\underline{0.9131} \\
    SDAN  & ours  &       & 410 K & \textbf{32.54}/\textbf{0.8989} & \textbf{28.81}/\textbf{0.7866} & \textbf{27.74}/\textbf{0.7413} & \textbf{26.65}/\textbf{0.8031} & \textbf{31.13}/\textbf{0.9152} \\
    \bottomrule
    \end{tabular}%
  \label{tab_4}%
\end{table*}%

\begin{figure*}[!t]
\centering
\includegraphics[width=6in]{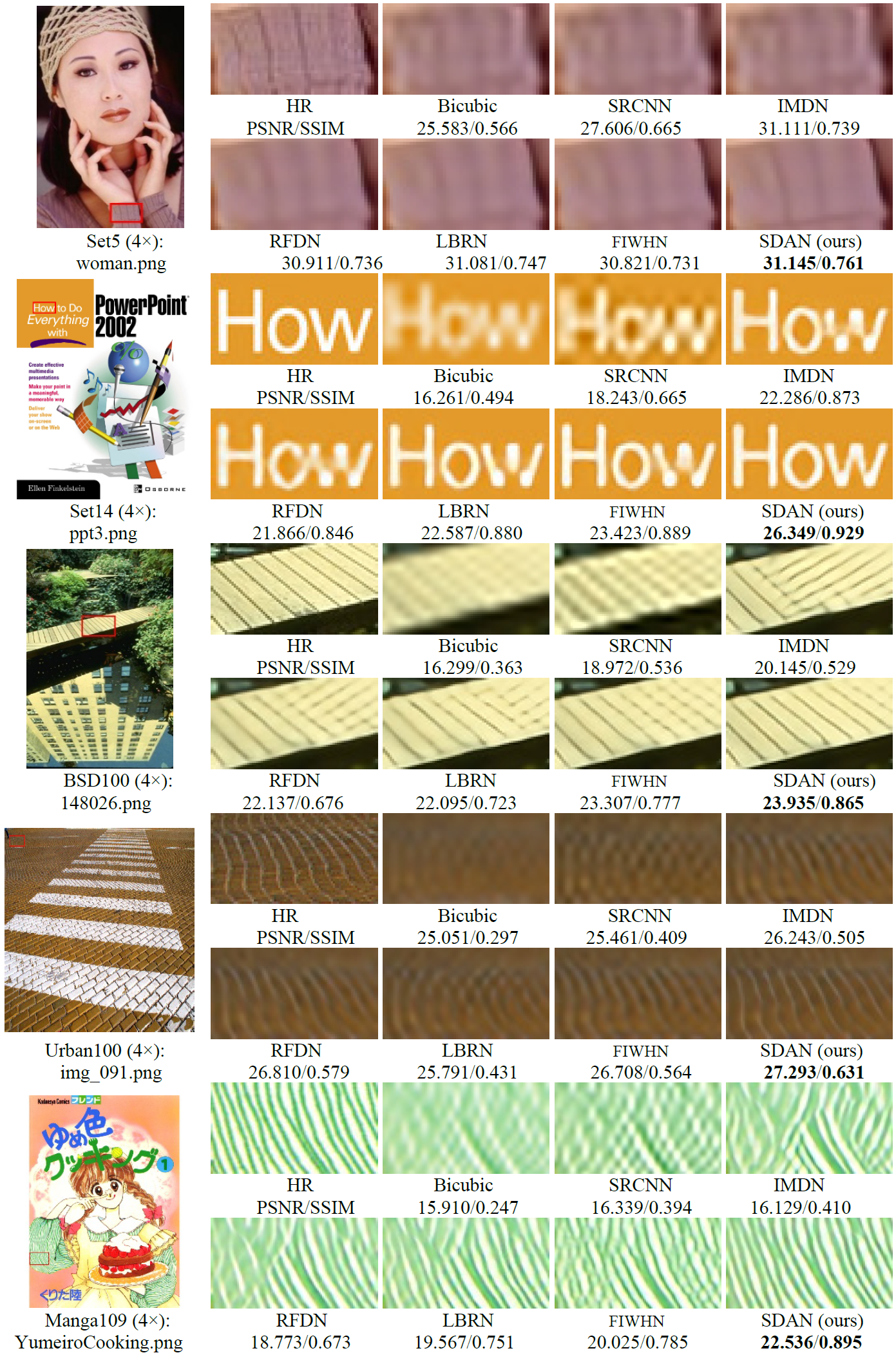}
\caption{Visual comparisons for ×4 SR with the BI model on the Set5, Set14, B100, Urban100 and Manga109 datasets. The best results are \textbf{highlighted}.}
\label{fig_5}
\end{figure*}

\textbf{PSNR/SSIM results.} The quantitative evaluation results for ×2, ×3 and ×4 lightweight SR are shown in Table \ref{tab_5}. For ×2 SR, in addition to achieving the majority of the top PSNR and SSIM values on five benchmark datasets, our SDAN has a minimum of 405 K model parameters compared with recent SR methods (PFFN, CFDN, FIWHN, and OSFFNet), demonstrating effectiveness of our SDAN. Using the same five benchmark datasets, our SDAN achieves the maximum PSNR and SSIM for ×3 SR with 407 K model parameters. Regarding ×4 SR, our SDAN achieves all of the best quantitative results of PSNR and SSIM. Compared to previous lightweight state-of-the-art methods, it is evident that our SDAN can attain superiority for nearly all scaling factors (e.g., ×2, ×3, and ×4). And, for larger amplification factors, the superiority of our method becomes more obvious. Overall, our method can produce superior reconstruction performance than other previous methods, as demonstrated quantitatively by the recorded experimental results. 
Furthermore, we compare the model complexity and performance of our SDAN for ×4 SR on the benchmark datasets with the heavyweight state-of-the-art SR methods in Table \ref{tab_4}, including EDSR \cite{ref51}, RDN \cite{ref7}, RCAN \cite{ref8}, DRN \cite{ref52}, and ERAN \cite{ref49}. On the Set14 and BSD100 datasets, our SDAN specifically achieves 28.81 dB and 27.72 dB respectively, and they are higher than corresponding values of the heavy network EDSR. Notably, the 43.1M parameters of EDSR are significantly more than the 0.41M parameters of our SDAN. Besides, it is noteworthy that our method obtains better performance than heavy network RDN on all four benchmark datasets (Set5, Set14, BSD100, and Urban100). Overall, our model requires at least 10 times less parameters than the heavier SR networks, yet it can still produce superior or comparable results, demonstrating its high efficiency. This attributes to the efficient information distillation via SDM module and the captured representative long-range dependencies by MM-LKA module.

\textbf{Visual results.} In Figure \ref{fig_5} , different methods are visually compared on selected images from the five benchmark datasets, i.e., SET5, Set14, BSDS100, URBAN100, and MANGA109. For "woman.png" image in Set5 dataset, the original HR image has a lot of texture and contour information.  The early bicubic algorithm produces the worst visual performance according to the these reconstruction results. In contrast, other CNN-based methods have produced better visual results. This trend also holds true for the quantitative results of PSNR and SSIM of these methods. For "ppt3.png" image in Set14 , the early bicubic approach is still not a good reconstruction algorithm for SR because of its unstable trend, aliasing artifacts, and extensive blurring, while the visual results of other CNN-based techniques (such as RFDN, LBRN, and FIWHN) have significantly been improved. It can be seen that our SDAN can achieve sharper details and more complete outlines, especially for English words in the image. Regarding "148026.png" image in BSDS100 dataset, all of these methods produce shape distortions, yet the result of our method has minimal shape distortions, being more natural and clearer. In terms of "148026.png" in URBAN100 and “YumeiroCooking.png” in MANGA109, our SDAN yields superior results with sharper details and more natural textures, and the objects in the reconstructed images have the most complete shapes and contours. Overall, our SDAN can achieve the highest PSNR and SSIM for test images due to the extracted discriminative feature with effective information distillation and multi-shape multi-scale contextual information, and it produces the best visual perception among all these methods by recovering sharper details and more complete contours. All these results visually and quantitatively show the effectiveness and superiority of the proposed SDAN.

\begin{table*}[htbp]
  \centering
  \caption{Performance and model complexity comparison of different lightweight SR methods (×2 Urban100). The best results are \textbf{highlighted}.}
    \begin{tabular}{p{4.045em}p{4.045em}p{4.045em}p{4.045em}p{4.5em}p{4.565em}p{3.5em}}
    \toprule
    Metric & \multicolumn{1}{p{4.045em}}{IMDN } & \multicolumn{1}{p{4.045em}}{RFDN} & \multicolumn{1}{p{4.5em}}{CFIN } & \multicolumn{1}{p{4.5em}}{FIWHN} & \multicolumn{1}{p{4.5em}}{OSFFNet } & \multicolumn{1}{p{3.5em}}{SDAN } \\
    \midrule
    {Years} & \multicolumn{1}{p{4.045em}}{MM19} & \multicolumn{1}{p{4.045em}}{ECCV20} & \multicolumn{1}{p{4.5em}}{TMM23} & \multicolumn{1}{p{4.5em}}{TMM24} & \multicolumn{1}{p{4.5em}}{{AAAI24}} & \multicolumn{1}{p{3.5em}}{ours} \\
    {Paras/K} & {694} & {534} & {675} & {705} & {516} & \textbf{405} \\
    {FLOPs/G} & {158.9} & {95.5} & {116.9} & {137.7} & \textbf{83.2}  & {93.1} \\
    PSNR/dB & {32.17} & {32.12} & {32.48} & {32.75} & {32.67} & \textbf{32.82} \\
    \bottomrule
    \end{tabular}%
  \label{tab_6}%
\end{table*}%

\begin{figure}[!t]
\centering
\includegraphics[width=2.75in]{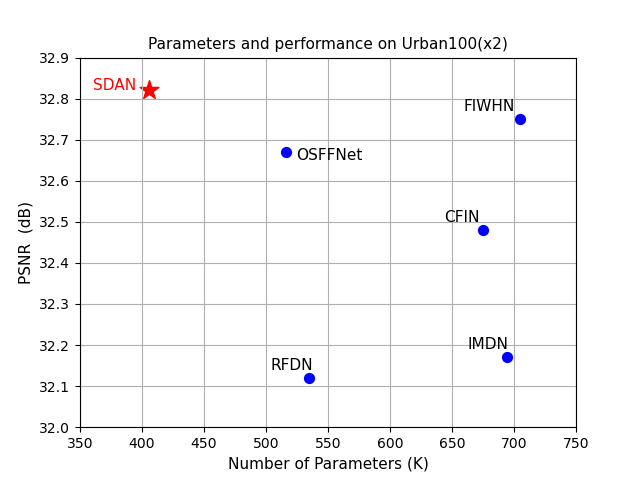}
\caption{Performance and the parameters of various lightweight SR methods on Urban100 (×2 SR).}
\label{fig_6}
\end{figure}

\begin{figure}[!t]
\centering
\includegraphics[width=2.75in]{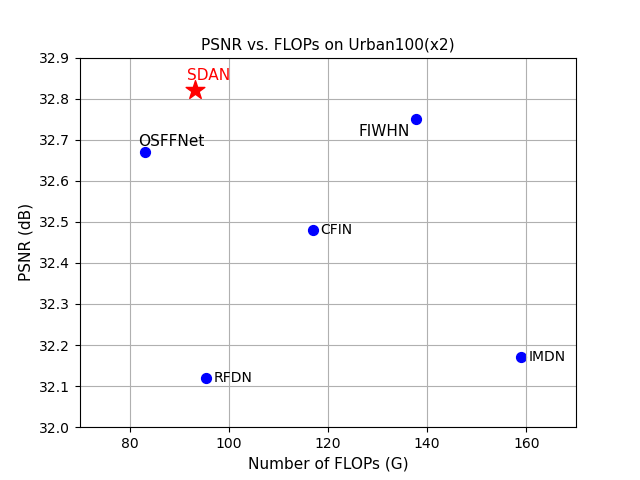}
\caption{Performance and the FLOPs of various lightweight SR methods on Urban100 (×2 SR).}
\label{fig_7}
\end{figure}

\subsection{Model Complexity Analysis}
On the Urban100 dataset for ×2 SR, we further analyze and record the model complexity and reconstruction performance of different methods including IMDN \cite{ref12}, RFDN \cite{ref13}, CFIN \cite{ref70}, FIWHN \cite{ref16}, OSFFNet \cite{ref18}, and our SDAN. Specifically, we adopt the widely used evaluation metrics, model parameters and FLOPs \cite{ref46} to show the model complexity, and we assume the SR output size to 1280× 720 to calculate FLOPs of these networks. Besides, we utilize the PSNR value to quantitatively evaluate the reconstruction quality of different outputs. The detailed results are shown in Table \ref{tab_6}. Among these methods, FIWHN possesses the most parameters 705 K while achieving the second highest PSNR 32.75 dB. By contrast, our SDAN with the fewest parameters 405 K can obtain the highest PSNR 32.82 dB. As for the FLOPs, OSFFNet has the fewest value 83.2 G and achieves 32.67 dB PSNR which is 0.15 dB lower than the 32.82 dB of our SDAN with 93.1 G FLOPs. These comparison results quantitatively show that our method is more efficient when dealing with image resolution degradation. The comparison of parameters and PSNR of these methods is visually illustrated in Figure \ref{fig_6}, and we further show their comparison of FLOPs and PSNR in Figure \ref{fig_7}.

\subsection{Limitations}
As shown in Figure \ref{fig_5}, although our model obtains better visual results than our competitors in terms of lightweight single-image super-resolution, there is still a significant gap between our results and label HR images. The performance of the algorithm and the richness of the datasets should be further improved in future research.  
\section{CONCLUSION}
In this study, we propose an efficient Star Distillation Attention Network (SDAN), a lightweight yet powerful framework for single-image super-resolution (SISR). The core innovation lies in the proposed Star Distillation Module (SDM) and the Multi-Shape Multi-Scale Large Kernel Attention (MM-LKA). SDM enhances discriminative representation learning via information distillation in the High-Dimensional Non-Linear (HDNL) feature spaces, and MM-LKA can capture multi-shape multi-scale contextual information effectively. By integrating SDM and MM-LKA, we further develop the Residual Star Distillation Attention Module (RSDAM) as the fundamental building block of SDAN, enabling hierarchical feature refinement while maintaining computational efficiency. Extensive experiments show the superiority of our method, demonstrating that SDAN with low model complexity achieves superior performance in both quantitative metrics and visual quality.

\section*{Acknowledgments}
The authors would like to thank the reviewers for their insightful comments and useful suggestions.



 
\bibliography{cas-refs.bib}


\newpage

 




\vfill

\end{document}